\documentstyle[12pt]{article}

\newcommand \al {\alpha}
\newcommand \bt {\beta}

\newcommand \de {\delta}

\newcommand \pr {\prime}
\newcommand \tdpr {\bar t}
\newcommand \tm {\tilde \mu}
\newcommand \tilt {\tau _J}
\newcommand \tb {\beta _J}
\newcommand \pdt {\partial _{t}}
\newcommand \si {\sigma}

\begin{document}

\topmargin 0pt
\oddsidemargin 5mm

\setcounter{page}{1}
\vspace{2cm}
\begin{center}

{\large \bf Spin-glass model with partially annealed asymmetric bonds.
 }\\

\quad

A.E. Allahverdyan, D.B. Saakian\\
{\em Yerevan Physics Institute}\\
{\em Alikhanian Brothers St.2, Yerevan 375036, Armenia}\\

\end{center}

\centerline{{\bf{Abstract}}}
We have considered the two-spin interaction spherical spin-glass model with
asymmetric bonds (coupling constants).
Besides the usual interactions between spins and bonds and between the 
spins and
a thermostat with temperature $T_{\sigma }$ there is also an additional 
factor:
the bonds are not assumed random {\it a priori} but interact with some 
other thermostat at the 
temperature $T_{J}$.
We show that when the bonds are frozen with respect to the spins a first 
order phase transition to a
spin-glass phase occurs,
and the temperature of this transition tends to zero if $T_J$ is large.
Our analytical  results show that a spin-glass phase can exist in
mean-field models with nonrelaxational dynamics.

\newpage
\vspace{3mm}

\noindent
\       \ Investigation of nonequilibrium stationary states for systems 
without detailed balance is
an important problem. Systems of this type are widely used in 
nonequilibrium thermodynamics
\cite{strat, shlogl}, biology and biophysics \cite{gorshkov}, theory 
of artificial neural networks \cite{parisi}, etc. \cite{noneqrev}.

If detailed balance (or potential conditions) holds, then the stationary 
state of a macroscopic system
can be described by the Gibbs
distribution (maybe with some generalizations),
which is {\it independent} from details of the dynamics \cite{gardiner, 
tomas}. In the opposite case 
such a universal and simple distribution does not exist, and the 
situation is strongly dependent on
the details of dynamics and the concrete 
form of detailed balance violation.
In general, we have in a stationary state without detailed balance only 
time-translation invariance
\cite{gardiner, tomas}:  average multitime quantities $C(t_1,..,t_n)$ 
(time ordering is assumed
$t_1\geq...\geq t_n$) depend only on the corresponding time differences 
$t_k-t_l$, $k< l$
(for example, $C(t)={\rm const}$, $C(t_1,t_2)=C(t_1-t_2)$,...)

In the present paper we consider the case when the detailed balance 
condition is violated
by nonconservative forces of some specific type: the mean-field spherical 
spin-glass model 
with asymmetric bonds (coupling constants). This type of nonconservative 
forces was inspired
by applications in the theory of artificial neural networks 
\cite{parisi}, but it has also some 
independent meaning as the model for a open spin-glass system.
The model was introduced in \cite{crisanti}, where was shown that
random frozen uncorrelated asymmetric gaussian bonds
(even when the asymmetry is small but generic) totally break 
finite-temperature spin-glass transition.
The second important step in this direction was done in \cite{kurchan}.
Here multispin interaction spherical spin-glass was considered,
and by numerical methods was shown that indeed the Langevin dynamics,
which is started from the typical 
initial conditions, tends to the paramagnetic  state for any nonzero 
temperature.
But if the initial conditions are chosen in some special way, then  
finite-temperature spin-glass
phase transition  to a metastable state is possible. Gugliandolo et. al. 
\cite{kurchan} proposed that this difference
between two-spin and multispin interaction models arises due to the
different structure of the phase spaces   when detailed balance 
holds (i.e., the bonds are symmetric). Indeed in the two-spin
interaction model the spin-glass phase is marginally stable, hence there 
is a possibility
to destroy the phase transition. In contrast, the multispin interaction 
model has low-lying
totally stable states \cite{crisanti1,crisanti2,aging}.

In  our model the coupling constants interact with the spins and with a 
thermal bath
at the temperature $T_J$. Generally speaking, can be $T_J\not= 
T_{\sigma}(\equiv T)$. 
There are several reasons for introducing  interactions of this type. A 
typical example of a 
two-temperature system is a nonequilibrium
electron-proton plasma. Due to the
large difference between electron and proton masses
energy transfer between the two components can be neglected in some range 
of times.
Protons and electrons then go to equilibrium independently and can have 
different temperatures.
In general, a nonequilibrium many-component systems
with slow energy transfer between the components can, in some range of 
times,  be described 
by introducing several temperatures.
Another reason is that: It is well known that in some cases and in some 
sense neural 
networks can be described by spin-glass
models. The spins and coupling constants are identified as neurons and 
synaptic connections.
But from a physiological point of view the nonlinear interactions with 
the neurons,
and varying the synaptic connections  with external environment
is a very important part of  recognition and memory.

At  first,  spin-glass models with slow varying bonds were considered by 
Horner \cite{horner}
\footnote{We thank referee for pointing these references.}. Here the 
model with the detailed balance condition
is considered, and the slow motion of bonds is used as some 
method  for obtaining correct long-time limits from dynamical equations. 
Other methods for this purpose are also 
developed by Horner and co-workers \cite{crisanti2}.

The Langevin equations for the model have the following form:
\begin{equation}
\label{2}
\tau \pdt \si _{i}=-r\si _{i}-\gamma \beta \sum _jJ_{ij}\si _j
 +\eta _{i}(t),\  \ \langle \eta _{i}(t)\eta
_{j}(t^{\prime })\rangle =2\tau \delta _{ij}\delta (t-t^{\pr }),
\end{equation}
\begin{equation}
\label{3}
\tilt \pdt J_{ij}
=\gamma \tb \si _i\si _j
-2v\tb J_{ij}+\eta _{ij}(t),
\  \
\langle \eta _{ij
}(t)\eta _{i^{\prime }j^{\prime }}(t^{\prime })\rangle =2\tilt \delta 
_{ii^{\prime }}\de _{jj^{\prime }}
\delta (t-t^{\pr }),
\end{equation}
where the bonds (spins) interact with the thermal bath at  temperature 
$T_J$ ($T_{\sigma }$),
$\gamma =\sqrt{2/N}$ is the standard norm-factor for mean-field models,
$r$ is the lagrangian factor for the spherical constraint $\sum _i\si ^2 
_i=N$ 
(here $N$ is the number of spins), 
$J_{ij}\not =J_{ji}$, and
$vJ_{ij}^2 $ is the potential energy for a coupling constant.

If the bonds are symmetric {\it a priori}, then 
Eqs. (\ref{2}) and (\ref{3}) can be viewed as equations generated by the  
mean-field Hamiltonian 
$$
H=\gamma \sum_{ij}J _{ij}\si _{i}\si _{j}
+v\sum _{ij}J^2_{ij}
$$ 
It is the Langevin dynamics of the well-known two-spin spherical model 
where the (symmetric) bonds
interact with the thermal bath at the temperature $T_J$ \cite{allah}.
In model considered in this paper
asymmetric and symmetric bonds have  the similar  potential energy,
and are subjected to the different thermal histories of the same thermostat.

The analysis of Eqs. (\ref{2}) and (\ref{3}) simplifies considerably in 
the 
thermodynamic limit when $N\to \infty$, where the dynamics
of the system can be described by a self-consistent equation involving a 
single spin only. This is achieved by introducing
generating functional for the Langevin dynamics \cite{dynfunc}. Because 
our interest is 
in the dynamics of the spins only, the
integration by the bonds can be taken in the generating functional.

The resulting {\it mean-field } equations read
\begin{eqnarray}
\label{4}
& &(\tau \pdt +r)\si (t)=\tm
\int_{-\infty}^{t} d\tdpr e^{-\al (t-\tdpr )} C(\tdpr ,t)\si (\tdpr
)
+\eta (t), \nonumber \\
& &
\langle \eta (t)\eta (t^{\prime })\rangle =2\tau \delta (t-t^{\pr })+\mu
\exp{(-\al |t-t^{\pr }|)} C(t,t^{\pr }),
\end{eqnarray}
where
$\al =2v\tb /\tilt $, $\tm
=2\bt \tb /\tilt $,
$\mu =\bt ^2/v\tb $. 
Our purpose here is to get equations for the correlation function
$C(t,t ^{\prime })= \langle \si (t)\si (t^{\pr }) \rangle$
and the response function
$G(t,t^{\pr })=\langle
\partial \si (t) / \partial \beta h (t^{\pr }) \rangle $. Because the 
initial time $\to -\infty $ (it is one of the limits of integration
in  Eq. (\ref{4})) the {\it equilibrium regime} is expected
$$
C(t,t^{\pr })=C(t-t^{\pr }),  \  \
G(t,t^{\pr })=G(t-t^{\pr }).
$$
It should be stressed again that the initial time tends to minus  
infinity {\it after} the thermodynamic limit. The resulting equations are
\begin{equation}
\label{5}
(\tau \pdt +r)C(t)=\tm \int_{-\infty}^{t}
d\tdpr e^{-\al (t-\tdpr )} C(t-\tdpr )C(\tdpr )
 +\mu \int_{0}^{\infty } d\tdpr  e^{-\al (t+\tdpr )}C(t+\tdpr)G(\tdpr )
\end{equation}
\begin{equation}
\label{6}
(\tau \pdt +r)G(t)=\tm \int_{0}^{t}
d\tdpr e^{-\al (t-\tdpr )} C(t-\tdpr )G(\tdpr )
\end{equation}
Here the difference of the times $t-t^{\pr }$ is denoted by the same 
letter $t$.
We also  take $\tau =1$ as fixation of the units.

Further, we shall investigate the {\it adiabatic limit}, where the bonds 
are frozen with respect of the dynamics of spins
\cite{gardiner, allah}. Indeed, in a qualitative manner we feel that any 
spin-glass order is possible only 
if bonds are frozen. This intuitively obvious fact has been rederived 
rigorously recently for spin-glass systems when
the detailed balance condition holds
\cite{horner,allah}. Thus the limit
\begin{equation}
\label{dopdop}
\tau _J \gg t-t^{\prime }
\end{equation} is assumed: The bonds are frozen when the spins are 
observed. Actually, this assumption is the standard
one for investigating the dynamics of usual spin-glass models when the 
bonds are frozen {\it a priori}.
The different factor of our approach is the noninfinite temperature 
$T_J$, which is a physical mechanism for introducing
a correlation between bonds. 

Now if $t-t^{\pr }$ is large enough, but stills much smaller than 
$\tau _J$ and all these times are much smaller than modulo of the initial 
time
which $\to -\infty$ just after $N\to \infty$, 
we expect  that the correlation function tends to the Edwards-Anderson 
order parameter
$$
C(t-t^{\prime } \mapsto \infty)=q.
$$
By the adiabatic condition (\ref{dopdop}) in  the last integral of Eq. 
(\ref{5})
we can take $e^{-\al \tdpr }\sim 1$ by considering the relevant domain of 
the integration.
In the first integral we have by interchanging variable
\begin{equation}
\label{bb1}
\tm
\int_{0}^{\infty }
d\tdpr e^{-\al \tdpr }
C^{p-1}(\tdpr )C(t-\tdpr )\mapsto \frac{\bt q^2}{v}.
\end{equation}

Similar analysis shows that the right-hand term in Eq. (\ref{6}) can be 
neglected,
and we have the simple solution
\begin{equation}
\label{moso}
G(t)=e^{-rt}
\end{equation}
There are three relevant parameters in our model: $v$, $\beta $, $\beta 
_J$. In this model 
we consider the phase diagram at the following conditions: $v=T_J$, 
$n=T/T_J$ is fixed, and
only $\beta =1/T$ is varying. Thus by 
\begin{equation}
\label{add1}
n\mapsto 0 
\end{equation}
we go to the
model was considered in  \cite{crisanti}.
\footnote{Furthermore, if the bonds in (\ref{2}, \ref{3}) are symmetric
{\it a priori} then by (\ref{dopdop}, \ref{add1}) we go to the usual 
spherical 
mean-field model \cite{allah}. Here $n$ plays the role of "replica 
number" which  in other 
approaches is introduced by the replica trick.} This fact can be derived 
directly from (\ref{3}): 
By limits (\ref{dopdop}, \ref{add1})
the bonds can be considered as uncorrelated, quench, asymmetric gaussian 
variables.

After this we have the following equation for $C(t)$:
\begin{equation}
\label{moso1}
(\pdt +r)C(t)=n\bt ^2q^2
 +\bt ^2 \int_{0}^{\infty } d\tdpr C(t+\tdpr)e^{-r\tdpr }
\end{equation}
This equation holds for $t\geq 0$.

Equation (\ref{moso1}) can be solved by the Laplace transformation.
As the general solution we have
\begin{equation}
\label{hez1}
C(t)=\frac{n\bt ^2q^2r}{r^2-\bt ^2}+k_{+}e^{t\sqrt{r^2-\bt ^2}}
+k_{-}e^{-t\sqrt{r^2-\bt ^2}} .
\end{equation}
The constants $k_{+}$, $k_{-}$, $r$ should be fixed by the following 
standard conditions:
$C(0)=1$ (the spherical constraint), $\pdt C(t)|_{t=0}=-1$, $k_{+}=0$ 
(the condition for monotonic decay of $C(t)$).
After some calculations we have for $r$ and Edwards-Anderson parameter $q$
\begin{equation}
\label{hez2}
q=\frac{n\bt ^2q^2r}{r^2-\bt ^2}
\end{equation}

\begin{equation}
\label{hez3}
r=\sqrt{\bt^2+\frac{1}{(1-q)^2}} .
\end{equation}
Finally we have the following equation for $q$:
\begin{equation}
\label{hez4}
\frac{q}{1-q}=n\bt ^2q^2\sqrt{\bt ^2(1-q)^2+1}
\end{equation}
Besides paramagnetic solution $q=0$ there is also the first-order phase 
transition into
a spin-glass phase (it is obvious that second-order transition is impossible
in this case). The transition point is defined as the first temperature 
when (\ref{hez4}) predicts 
a non-zero (non-paramagnet) solution. 
The temperature and the jumping of $q$ at this transition point can be 
obtained by the following equations:
\begin{equation}
\label{hez44}
q_c=\frac{\psi +1}{3\psi +2} ,
\end{equation}
\begin{equation}
\label{hez5}
T_c=\frac{1-q_c}{\sqrt{\psi}} ,
\end{equation}
where $\psi $ is the positive solution of the equation
\begin{equation}
\label{hez6}
2\psi +1=n\psi (1+\psi )^{3/2} .
\end{equation}
When $n\mapsto 0$ (in this limit we go to the particular case that was 
studied 
in Ref. \cite{crisanti}) we have $T_c\mapsto 0$, $q_c\mapsto 1/3$.

Further, we should discuss the stability properties of the spin-glass 
solution. If detailed balance holds, then
static properties of a system is described by the Gibbs distribution 
therefore the stability of solutions can be investigated by
analyzing minimums of free energy \footnote{A remark should be given 
here: Because the mean-field dynamical equations
are obtained by $N\to \infty$ before the initial time $\to -\infty$, some 
states that arise in dynamics cannot be reflected in
the purely static investigation by means of free energy. Thus,  phase 
transitions that are predicted by statics and dynamics 
can be different \cite{crisanti2,aging}}. 
But for this concrete problem free energy is not necessary: Unstability 
of a solution
is reflected by unstable 
behavior of purely dynamical quantities such as positive derivation of 
correlation function, 
or negative susceptibility. For example, the famous (AT) line in usual 
spin-glass models can be recovered by this purely
dynamical consideration \cite{crisanti2,som}. 
This property does not connect with any specific character of spin-glass 
systems but is 
supported by general theorems about the stability 
of stochastic dynamical systems (generally speaking without detailed 
balance) \cite{strat,shlogl,tomas}.
Because in our case there are no such anomalies both for paramagnet and 
spin-glass
solutions we conclude that these phases are stable (by chosen initial 
conditions). 
The stability of paramagnet for any temperature is more or less typical for
first-order phase transitions \cite{crisanti1,aging,allah}. 

Now about an other  important problem: If we have two different stable
phase, then one should be only metastable and another one true stable. In 
usual models this question is investigated by free energy:
For the true stable phase the free energy should be minimal. In 
principle, a generalized free energy can be introduced also in 
models without detailed balance; the corresponding quantity is based on 
the so called Kullbak entropy or relative entropy 
\cite{strat,shlogl}.
For systems with non-gibbsian stationary distribution this function plays 
nearly the same role as usual free energy
but, generally speaking, has a strong dependence from initial conditions 
| in the usual case the detailed balance 
condition removes this dependence.
Unfortunately,  we have not succeeded in calculation of this function for 
our model.
So really we found only that  the spin-glass phase occur at least as a 
metastable state. On other hand all 
stable spin-glass solutions | known by other models (see, for example, 
\cite{crisanti2, allah, gross}) |  are true
stable with respect to paramagnet at least if temperature is low enough 
(of course it is not an argument for true stability in our case
but rather a hint). One of the scenarios for phase transitions realized 
in such systems is as follows \cite{allah, nagaev}: 
At some temperature the nontrivial phase (spin-glass in our case) occurs 
at the first time,
but true thermodynamical transition occurs at some lower temperature 
where free-energies of the
different solutions are the same. May be the scenario for our phase 
transition is the same.

In the low-temperature limit we have for our solution
\begin{equation}
\label{hez7}
1-q\sim \frac{T^2}{n}
\end{equation}
We get this equation by the assumption that $1-q$ is small.
It should be noted that the characteristic time for the correlation 
function relaxation is                    
$t_{rel}=1-q$. Thus, critical slowing down occur only near 
$T=0$.                                                          

Now about the origin of this spin-glass phase. As we mentioned above
behavior of a system without detailed balance can be connected with behavior
of the phase space of the same system when detailed balance holds (i.e., 
the bonds are 
symmetric in our case). 
If we assume that such arguments can be used, then 
at this qualitative level the situation is clear: as was shown in 
\cite{allah} a
finite $n$ breaks marginality of the spin-glass solution in the two-spin 
spherical model.
'Replicon' (or dangerous) eigenvalue is $\sim n$ (except exactly the 
critical point). 
Thus, at least, some spin-glass states are totally stable, and can
'struggle' against the nonhamiltonian influence.

We consider the spherical spin-glass model with asymmetric partially 
annealed bonds and show that this last property can 
induce a transition to the spin-glass phase. 
Several {\it very} important problems in this field should be considered 
in future: The problems of proving
true stability for the new phase, considering partially asymmetric bonds, 
and of course investigating of more realistic models. 

{\bf Acknowledgments.}

Authors thank unknown referee for physically very well-motivated remarks. 
A.E. Allahverdyan also thank
E. Mamasakhlisov for helpful discussions.

\end{document}